\documentclass[%
 reprint,
 amsmath,amssymb,
 aps,
]{revtex4-2}
\usepackage{subcaption}
\usepackage{graphicx}
\usepackage{dcolumn}
\usepackage{bm}
\usepackage{braket}

\def\leqn#1{(\ref{#1})}
\usepackage{float}

\def\beq{\begin{equation}}
\def\eeq#1{\label{#1}\end{equation}}
\def\eeqn{\end{equation}}

\newenvironment{Eqnarray}%
   {\arraycolsep 0.14em\begin{eqnarray}}{\end{eqnarray}}
\def\beqa{\begin{Eqnarray}}
\def\eeqa#1{\label{#1}\end{Eqnarray}}
\def\eeqan{\end{Eqnarray}}

\def\leqn#1{(\ref{#1})}

\def\ee{e^+e^-}

\begin{document}

\title{Sensitivity of Heavy Higgs Boson to the Precision Yukawa
  Coupling Measurements at Higgs Factories}

\author{Kamal Maayergi}
\affiliation{Department of Physics and Astronomy, Dartmouth College \\Hanover, NH 03755 USA}%

\author{Devin G.~E.~Walker}
\affiliation{%
 Department of Physics and Astronomy, Dartmouth College \\Hanover, NH 03755 USA
}%

\author{Michael E.~Peskin}
\affiliation{
SLAC National Accelerator Laboratory\\ Stanford University\\ Menlo Park, California 94025 USA
}%

\begin{abstract}
We investigate the potential of precision Higgs boson coupling measurements to discover heavy Higgs bosons by performing scans of the parameter space in Two-Higgs-Doublet Models (2HDM). Our study encompasses conventional Type I and Type II models, as well as models in which Higgs couplings differ between the third generation and lighter fermion generations. The scans reveal that precision measurements at the sensitivity levels projected for Higgs factories, such as Linear Collider Facility~(LCF) and the FCC-ee at CERN and the CEPC in China, are capable of probing heavy Higgs boson masses in the multi-TeV range, with sensitivity extending beyond 5 TeV in certain scenarios. In particular, the precise determination of the charm quark Yukawa coupling at Higgs factories provides a powerful test of the hypothesis that the fermion mass hierarchy arises from an extended Higgs sector with different  Higgs fields coupling to the different generations of fermions.

\end{abstract}

\maketitle

\section{\label{sec:Intro} Introduction}

Since the discovery of the Higgs boson in 
2012~\cite{ATLAS:2012yve,CMS:2012qbp}, the ATLAS and CMS experiments 
at the LHC have extensively probed the couplings of this particle to 
vector bosons and fermions~\cite{CMS:2018uag,ATLAS:2019nkf}. Results to date have all been 
consistent with the predictions for the unique Higgs boson of the 
Standard Model (SM), with coupling measurements achieving accuracies at 
the 10\%–20\% level for the $W$ and $Z$ bosons and the third-generation 
quarks and leptons. This is a tremendous achievement, firmly establishing 
the 125~GeV Higgs boson as the principal source of spontaneous 
electroweak symmetry breaking. The upcoming High-Luminosity run of 
the Large Hadron Collider (HL-LHC) will improve these results in two key ways: (1) by increasing the precision of coupling measurements to the 1–4\% level; and (2) by extending the search for additional scalar bosons in the Higgs sector to beyond the 1~TeV mass scale~\cite{ATLAS:2022hsp}. In 
this paper, we refer to the known Higgs boson as $h$ and a potential 
heavier boson partner as $H$.  

Nonetheless, our understanding of the nature of the Higgs boson and the mechanism of electroweak symmetry breaking remains incomplete.  In the SM, electroweak symmetry breaking is an assumption, implemented by the choice of input parameters.  Moreover, the SM does not explain the large hierarchy of fermion masses,  with ratios exceeding 100,000 between the heaviest and lightest fermions.  It is therefore crucial to obtain new insights into the properties and 
possible partners of the 125~GeV Higgs boson. This motivation has led to 
proposals for new $\ee$ colliders designed as ``Higgs factories'' to 
enable higher-precision measurements of the known Higgs boson $h$~\cite{Dawson:2022zbb}. 
Current proposals under consideration include the ILC~\cite{ILCInternationalDevelopmentTeam:2022izu} 
in Japan, the CEPC in China~\cite{CEPCPhysicsStudyGroup:2022uwl}, and the FCC-ee 
and LCF at CERN~\cite{FCC:2025lpp,LinearColliderVision:2025hlt}. The projected 
precision for Higgs boson coupling measurements at these facilities is 
quite similar; in this paper, we refer to the LCF projections for definiteness.

A question that is often asked about precision measurements is: What
is the mass reach for the sensitivity to new particles?  A simple
estimate of the sensitivity can be derived from the idea that
corrections to the SM due to new heavy particles
should be parametrized by higher-dimension operators
in Standard Model Effective Field Theory (SMEFT).   The leading such
operators are at dimension~6, leading to an estimate of the size of
the corrections as
\beq
v^2/M^2 \sim  1\% \ ,
\eeq{corrval}
where $v$ is the Higgs field vacuum expectation value and $M$ is the
heavy particle mass, possibly suppressed further by a factor of
$\alpha_w$.  However, \leqn{corrval} is only an order-of-magnitude
estimate which might be modified by a large dimensionless prefactor.
There are certainly beyond-SM theories that give small corrections to
the Higgs couplings.  A more meaningful question  is whether there are opportunities
for discovery with Higgs factory precision measurements~\cite{Peskin:2022pfv}.
To address this, it is necessary to study explicit models of
new physics.  It is well-known that composite models of the Higgs
boson such as the Strongly-Interacting Light Higgs
(SILH)~\cite{Giudice:2007fh} can create observable deviations in the
Higgs couplings from mass scales of 10~TeV and above.   In this paper,
we will address whether this is possible in essentially weak-coupling
multi-Higgs models.

The most familiar class of such models is the two-Higgs-doublet-model
(2HDM)~\cite{Gunion:1989we}.  There are many types of 2HDM models,
depending on the pattern chosen for the Higgs boson couplings to
fermions.  The most well-studied is the Type II 2HDM, with the pattern
of couplings found in Minimal Supersymmetry.  In this model, the light
Higgs boson $h$ 
obeys minimal flavor violation (MFV), which means that flavor changing
neutral currents (FCNC) are suppressed by linking all
flavor and CP-violating interactions to the known Yukawa couplings of
the SM~\cite{Chivukula:1987py, DAmbrosio:2002vsn}.  Even here, as was
pointed out in \cite{Peskin:2022pfv}, the restriction to
supersymmetric models places a strong limit on the size of the
expected correction to Higgs couplings.   Here, we will give up the
assumption of supersymmetry and make a broader exploration of the parameter space.

Another model that, in our opinion, has not received sufficient
attention, is the ``Flavorful
2HDM''~model~\cite{Altmannshofer:2016zrn,Altmannshofer:2017uvs,Altmannshofer:2018bch}.
This is the simplest realization of the intuitive notion that the quark and lepton
flavor hierarchy is the result of having different Higgs bosons give
mass to the fermions of the various generations. The lightest Higgs
$h$ is a linear combination of the various more fundamental Higgs
fields~\cite{Ross:1989qk,Chan:1997cv,Bjorken:2002vt}.
This model gives up
MFV but, as shown in~\cite{Altmannshofer:2016zrn,Altmannshofer:2017uvs,Altmannshofer:2018bch}, the extra flavor violation
is well within experimental bounds.  In these models, the couplings of
$h$ to the third generation quarks and leptons can be close to the SM
values while the Yukawa couplings of first- and second-generation
quarks and leptons can be very different.   The opportunity at Higgs
factories to provide a measurement of the charm quark Yukawa coupling
at the percent level can provide an especially  strong test of this model and other
models of this general class.

The structure of this paper is as follows:   In Section~II, we will
review the general structure and parameter set of the 2HDM, and we
will review the structure of the conventional Type II 2HDM models. We also define some variants of this models that are useful for comparison. In
Section~III, we will review the structure of the Flavorful 2HDM and its similar variants.
Section~IV addresses the reach of these models in the commonly used parametrization, focusing on the mixing between the light and heavy Higgs bosons.  In Section~IV, we will present parameter scans of the sensitivity of Higgs Yukawa couplings to the 2HDM
physics, first for the conventional Type II models and then for the
Flavorful models, considering four different variants in each case.

\section{Structure of Two-Higgs-doublet models}
\label{sec:2HDM}

\subsection{Scalar Potential}
The Two-Higgs-doublet model (2HDM) is a well-studied extension of the Standard Model in which two Higgs doublets are present, unlike the SM Higgs sector that contains only one Higgs doublet~\cite{Gunion:1989we}.  In this section, we give a quick review of the 2HDM The most general gauge-invariant scalar potential is given by~\cite{Gunion:2002zf},
\begin{equation} 
\begin{split}
V& =m_{11}^2 \,\Phi_1^\dag\Phi_1 + m_{22}^2\, \Phi_2^\dag\Phi_2 - \left(\,m_{12}^2 \Phi_1^\dag \Phi_2 + \mathrm{h.c.}\,\right) \\ 
& +\frac{1}{2}\lambda_1(\Phi_1^\dag\Phi_1)^2 + \frac{1}{2}\lambda_2(\Phi_2^\dag\Phi_2)^2 + \lambda_3(\Phi_1^\dag\Phi_1)(\Phi_2^\dag\Phi_2)  \nonumber \\ 
&+ \lambda_4(\Phi_1^\dag\Phi_2)(\Phi_2^\dag\Phi_1) + \Bigl(\,\frac{1}{2}\lambda_5(\Phi_1^\dag\Phi_2)^2  \nonumber\\
 & + \left(\,\lambda_6(\Phi_1^\dag\Phi_1) + \lambda_7(\Phi_2^\dag\Phi_2)\right)\Phi_1^{\dag}\Phi_2 +\mathrm{h.c.}\,\Bigr), \nonumber
\end{split}
\label{Higgspot}
\end{equation}
where the terms $m_{12}^2$, $\lambda_5$, $\lambda_6$, and $\lambda_7$
can be complex. In this study we ignore the possibility of explicit CP
violating effects in the Higgs potential, by only taking the real
parts of the coefficients in equation \leqn{Higgspot}. We assume that
the parameters $\lambda_{1-7}$ are chosen such that the minimum of the
potential in equation \leqn{Higgspot} respects the $U(1)_\mathrm{em}$ gauge
symmetry~\cite{Gunion:2002zf}.  In this paper, we denotate the various Higgs boson mixing angles using
the shorthand notation $c_\beta \equiv \cos\beta$, $s_\beta \equiv
\sin\beta$, $c_\alpha \equiv \cos\alpha$, $s_\alpha \equiv
\sin\alpha$, $c_{2\alpha} \equiv \cos2\alpha$, $s_{2\alpha} \equiv
\sin2\alpha$, $c_{\beta-\alpha} \equiv \cos(\beta-\alpha)$,
$s_{\beta-\alpha} \equiv \sin(\beta-\alpha)$, and $t_\beta \equiv
\tan\beta$.

The scalar field vacuum
expectation values are then given by 
\begin{align} 
\braket{\Phi_1}=\frac{1}{\sqrt{2}}
\begin{pmatrix}
  0 \\
  v_1\\
\end{pmatrix}
&& 
\braket{\Phi_2}=\frac{1}{\sqrt{2}}
\begin{pmatrix}
  0 \\
  v_2\\
\end{pmatrix}
\end{align} 
where the terms $v_1$, and $v_2$ are taken to be real. The minimum of the potential requires~\cite{Gunion:2002zf}
\begin{align}\label{m11}
\begin{split}
&m_{11}^2 =m_{12}^2\,t_\beta\\
&\qquad-\frac{1}{2}v^2(\lambda_1\,c^2_\beta+\lambda_{345}\,s^2_\beta+3\lambda_6\,s_\beta \,c_\beta+\lambda_7\, s^2_\beta\, t_\beta)
\end{split}
\end{align}

\begin{align}\label{m22}
\begin{split}
&m_{22}^2 =m_{12}^2\,t^{-1}_\beta\\
&\qquad-\frac{1}{2}v^2(\lambda_2\,s^2_\beta+\lambda_{345}\,c^2_\beta+\lambda_6\,c^2_\beta \,t^{-1}_\beta+3\lambda_7\, s_\beta \,c_\beta) 
\end{split}
\end{align}
where we have defined the terms 
\begin{equation}\label{l345_tanb}
    \lambda_{345}=\lambda_3+\lambda_4+\lambda_5, \;\;\; \tan\beta=v_2/v_1,
\end{equation}
and the vacuum expectation value for the SM Higgs
\begin{equation}\label{vev}
    v^2=v_1^2+v_2^2=(246 \; \text{GeV})^2.
\end{equation}
It will be useful for us to define~\cite{Gunion:2002zf} a linear combination of the $\lambda_i$.
\begin{align}
    \lambda &\equiv \lambda _1\,c^4_{\beta }+\lambda _2\,s^4_{\beta }+\frac{1}{2}\lambda _{345}\,s^2_{2\beta } \label{lambda}  \\
    &\qquad+2s_{2\beta }\left(\lambda _6\,c^2_{\beta }+\lambda _7\,s^2_{\beta }\right) \nonumber \\ \nonumber \\
\hat{\lambda }&\equiv \frac{1}{2}s_{2\beta }\left(\,\lambda _1\,c^2_{\beta }-\lambda _2\,s^2_{\beta }-\lambda _{345}\,c_{2\beta }\,\right) \label{lambdahat} \\
    &\qquad-\lambda _6\,c_{\beta }\,c_{3\beta }-\lambda _7\,s_{\beta }\,s_{3\beta } \nonumber
\end{align}

In the decoupling limit where $\alpha \sim \beta - \pi/2$, we have~\cite{Gunion:2002zf} 
\begin{equation}\label{cbma}
    \text{cos}\left(\beta -\alpha \right)\approx \frac{\hat{\lambda}\, v^2}{m^2_H-m^2_h}
\end{equation}
where 
\begin{equation}\label{mh}
m_h^2\approx v^2\left(\lambda -\hat{\lambda }\,c_{\beta -\alpha }\right),    
\end{equation}
and
\begin{equation}\label{mH}
    m^2_H\approx v^2\left(\frac{\hat{\lambda }}{c_{\beta -\alpha }}+\lambda -\frac{1}{2}\hat{\lambda }\,c_{\beta -\alpha }\right)
\end{equation}
where equation \leqn{cbma} yields an $\mathcal{O}(v^2/m^2_A)$
correction to the value of $\cos(\beta-\alpha)$ as we approach the
decoupling limit. Then $\sin(\beta-\alpha) = 1 - \mathcal{O}(
v^4/m^4_A)$
and the corrections to the $h$ couplings to the $W$ and $Z$ bosons are
very small in this limit. 
Also, in the decoupling limit,  the masses of
the heavier Higgs
boson are approximately degenerate,  $m_H \approx m_A \approx
m_{H^\pm}$.   These features will appear in any 2HDM.  The ``alignment limit" defined by $\beta - \alpha = \pi/2$ recovers Standard Model-like couplings for the light Higgs $h$. Deviations from this limit result in modified couplings that can be probed by precision measurements at current and future colliders.

\subsection{Yukawa Sector}

The Yukawa sector in Type II 2HDM is
\begin{align}
\mathcal{L}_{\mathrm{Yukawa}} = &- Y_{ij}^d \, \overline{Q}_{Li} \Phi_1 d_{Rj}- Y_{ij}^u \, \overline{Q}_{Li} \widetilde{\Phi}_2 u_{Rj} \\
&- Y_{ij}^\ell \, \overline{L}_{Li} \Phi_1 e_{Rj} + \mathrm{h.c.} \nonumber
\end{align}
where $Q_{Li}$ and $L_{Li}$ are the left-handed quark and lepton doublets.  $d_{R}$, $u_{R}$, and $e_{R}$ are the right-handed down-type quark, up-type quark, and lepton singlets, respectively, $Y^{u}$, $Y^{d}$, and $Y^{\ell}$ are the corresponding Yukawa coupling matrices, where $i, j = 1, 2, 3$ are generation indices.  $\widetilde{\Phi}_{2} \equiv i \sigma_{2} \Phi_{2}^{*}$, where $\sigma_{2}$ is the second Pauli sigma matrix.  It is clear $\Phi_1$ couples exclusively to down-type quarks and charged leptons, while $\Phi_2$ couples to up-type quarks.   This coupling ensured by the following discrete $Z_2$ symmetry,
\begin{align}
    \Phi_1 &\to - \Phi_1 &  \Phi_2 &\to \Phi_2\, \\
    d_{Rj} &\to - d_{Rj} &  e_{Rj} &\to - e_{Rj}\,,
\end{align}
which is softly broken by $m_{12}$.  This discrete symmetry ensures an absence of tree-level flavor-changing neutral currents. 

Within the set of 2HDM models with minimal flavor violation, there are
a number of distinct choices for the couplings to fermions. We will
refer to models in this class as A models, and the corresponding
models described in the next section as B models.  We will
use the following notation:  We will differentiate $\Phi_1$ and
$\Phi_2$ in equation~\leqn{Higgspot}  by the condition $m_{11}^2 > m_{22}^2$, so
that generally $\tan\beta > 1$  in
\leqn{Higgspot}. In Type 1A models, all of the fermions
receive mass from the same Higgs field, which we take to be $\Phi_2$.
In Type 2A models, the leptons and down quarks couple to $\Phi_1$
while the up quarks couple to $\Phi_2$.   In the lepton-specific model
Type 2AL, the leptons couple to $\Phi_1$ while the quarks couple to
$\Phi_2$.  In the flipped model Type 2AF, the leptons and down quarks
couple to $\Phi_2$ while the up quarks couple to $\Phi_1$.   These assignments, and the corresponding assignments for the B models to be described in the next section, are summarized in Table~\ref{tab:assignments}.

\begin{table}
\begin{center}
\begin{tabular}{lcccccc}
Model     &   $u_{1,2}$  &   $u_3$  &  $d_{1,2}$ &  $d_3$   &
                                                              $e_{1,2}$
  &  $e_3$   \\
  \hline  \\
Type 1A  &     $ \Phi_2 $        &  $ \Phi_2 $     &   $ \Phi_2 $     &  $ \Phi_2 $    &  $ \Phi_2 $        &   $ \Phi_2 $    \\
Type 1B   &   $ \Phi_1 $        &  $ \Phi_2 $     &   $ \Phi_1 $     &  $ \Phi_2 $    &  $ \Phi_1 $        &   $ \Phi_2 $   \\ \\
Type 2A   &    $ \Phi_2 $        &  $ \Phi_2 $     &   $ \Phi_1 $     &  $ \Phi_1 $    &  $ \Phi_1 $        &   $ \Phi_1 $     \\
Type 2B    &    $ \Phi_1 $        &  $ \Phi_2 $     &   $ \Phi_2 $     &  $ \Phi_1$    &  $ \Phi_2 $        &   $ \Phi_1 $    \\ \\
Flipped A  &      $ \Phi_2 $        &  $ \Phi_2 $     &   $ \Phi_1 $     &  $ \Phi_1 $    &  $ \Phi_2 $        &   $ \Phi_2 $   \\
Flipped B   &   $ \Phi_1 $        &  $ \Phi_2 $     &   $ \Phi_2 $     &  $ \Phi_1 $    &  $ \Phi_1 $        &   $ \Phi_2 $    \\ \\
Lepton-specific A  &  $ \Phi_2 $   &  $ \Phi_2 $     &   $ \Phi_2 $     &  $ \Phi_2 $    &  $ \Phi_1 $        &   $ \Phi_1 $    \\
Lepton-specific B  &  $ \Phi_1 $        &  $ \Phi_2 $     &   $ \Phi_1 $     &  $ \Phi_2 $    &  $ \Phi_2 $        &   $ \Phi_1 $     \\ 
\end{tabular}
\end{center}
 \caption{Pattern of the couplings of the SM fermions to the two Higgs doublets $\Phi_1$ and $\Phi_2$ in each of the 2HDM models that we consider. In the models of type A, with natural flavor conservation, all examples each type of fermion $(u,d,e)$ couple to the same Higgs doublet.   In the flavorful models of type B, the third generation quarks and leptons couple to the opposite Higgs doublets from the first and second generation quarks and leptons.}
\label{tab:assignments} 
 \end{table}

\section{Flavorful Two-Higgs-Doublet models}
\label{sec:FlavorTextures}

The F2HDM is a setup in which one of the Higgs doublets is responsible
for the mass of the third generation SM fermions, while the second
Higgs doublet is responsible for the masses of the first and second
generation SM fermions~\cite{Altmannshofer:2016zrn}. In principle, the
smallness of the first and second generation fermion masses could be
the result of this structure, with the couplings of these fermions to
the $h$ suppressed by decoupling effects.  The analysis of this
decoupling generally leads to an enhancement of the first and second
generation Yukawa couplings relative to the SM predictions, and a
smaller suppression of the third generation Yukawa couplings.  The
F2HDM is a simple model that makes these considerations concrete.

To define the model, we consider the following Yukawa textures~\cite{Altmannshofer:2016zrn}
\begin{equation}\label{lepton yukawa}
    \lambda ^l\sim\:\frac{\sqrt{2}}{v_1}\begin{pmatrix}0&0&0\\ \:0&0&0\\ \:0&0&m_{\tau }\end{pmatrix}\:,\:\lambda '^l\sim\:\frac{\sqrt{2}}{v_2}\begin{pmatrix}m_e&m_e&m_e\\ m_e\:&m_{\mu \:}&m_{\mu \:}\\ \:m_{e\:}&m_{\mu \:}&m_{\mu \:}\end{pmatrix}
\end{equation}
\\this texture is chosen because it gives the observed lepton masses and it naturally explains the hierarchy between the second and third generation lepton masses. The  Yukawa texture for the down type quarks that naturally leads to the observed down-quark masses and CKM mixing angles is given by~\cite{Altmannshofer:2016zrn}

\begin{align}\label{down yukawa}
    \lambda ^d\sim\:\frac{\sqrt{2}}{v_1}\begin{pmatrix}0&0&0\\ \:0&0&0\\ \:0&0&m_b\end{pmatrix}\:,\:\lambda '^d\sim\:\frac{\sqrt{2}}{v_2}\begin{pmatrix}m_d&\lambda m_s&\lambda ^3m_b\\ m_d\:&m_{s\:}&\lambda ^2m_{b\:}\\ \:m_{d\:}&m_{s\:}&m_{s\:}\end{pmatrix}
\end{align}
where $\lambda \approx 0.23$, and is the Cabibbo angle. The third Yukawa texture is that of the up-type quarks which is analogous to that of the lepton sector~\cite{Altmannshofer:2016zrn}
\begin{equation}\label{up yukawa}
     \lambda ^u\sim\:\frac{\sqrt{2}}{v_1}\begin{pmatrix}0&0&0\\ \:0&0&0\\ \:0&0&m_{t}\end{pmatrix}\:,\:\lambda '^u\sim\:\frac{\sqrt{2}}{v_2}\begin{pmatrix}m_u&m_u&m_u\\ m_u\:&m_{c}&m_{c}\\ \:m_{u}&m_{c}&m_{c}\end{pmatrix}
\end{equation}
From these Yukawa textures we will get modifications to the Yukawa
couplings of the light Higgs to the SM particles. The equations of
interest in this study are the Yukawa couplings \cite{Altmannshofer:2016zrn}
\begin{equation}\label{Y_h}
    Y_\ell^h \equiv \langle \ell_L | Y_h^\ell | \ell_R \rangle = \frac{m_\ell}{v} \left( \frac{c_\alpha}{s_\beta} - \frac{m'_{\ell\ell}}{m_\ell }\frac{c_{\beta-\alpha}}{s_\beta c_\beta} \right)
\end{equation}
in the above equation, $\ell = e, \mu, \tau$. Then for
the flavor violating Higgs couplings
\begin{equation}\label{Y_hflav}
    Y_{\ell\ell'}^h \equiv \langle \ell_L | Y_h^{\ell} | \ell'_R \rangle = -\frac{m'_{\ell\ell'}}{v}\frac{c_{\beta-\alpha}}{s_\beta c_\beta}
\end{equation}
where in this case, $\ell \neq \ell'$. From these equations, we have
the modifications of the Yukawa couplings relative to the SM
predictions. There are two distinct sets of relations, those of the
first and second generation fermions, and those for the third
generation which are governed
by different doublets \cite{Altmannshofer:2016zrn}.  For the lighter generations,
\begin{equation}\label{kmu}
    \kappa _{\mu }\equiv \frac{Y_{\mu }}{Y_{\mu }^{SM}}=-\frac{s_{\alpha }}{c_{\beta }}+\mathcal{O}\left(\frac{m_{\mu }}{m_{\tau }}\right)\times \frac{t_{\beta }}{s_{\beta }^2}c_{\beta -\alpha },
\end{equation}
\begin{equation}\label{kc}
    \kappa _c\equiv \frac{Y_c}{Y_c^{SM}}=-\frac{s_{\alpha }}{c_{\beta }}+\mathcal{O}\left(\frac{m_c}{m_t}\right)\times \frac{t_{\beta }}{s_{\beta }^2}c_{\beta -\alpha },
\end{equation}
\begin{equation}\label{ks}
    \kappa _s\equiv \frac{Y_s}{Y_s^{SM}}=-\frac{s_{\alpha }}{c_{\beta
      }}+\mathcal{O}\left(\frac{m_s}{m_b}\right)\times \frac{t_{\beta
      }}{s_{\beta }^2}c_{\beta -\alpha }\ .
\end{equation}
These expressions hold equally well for the first generation fermions upon replacement of the second generation mass terms with the first generation masses \cite{Altmannshofer:2016zrn}. For the third generation fermions we have \cite{Altmannshofer:2016zrn}
\begin{equation}\label{kt}
\kappa_t \equiv \frac{Y_t}{Y_t^{\mathrm{SM}}} = \frac{c_\alpha}{s_\beta} + \mathcal{O}\left(\frac{m_c}{m_t}\right) \times \frac{t_\beta}{s_\beta^2} c_{\beta-\alpha},
\end{equation}

\begin{equation}\label{kb}
\kappa_b \equiv \frac{Y_b}{Y_b^{\mathrm{SM}}} = \frac{c_\alpha}{s_\beta} + \mathcal{O}\left(\frac{m_s}{m_b}\right) \times \frac{t_\beta}{s_\beta^2} c_{\beta-\alpha},
\end{equation}

\begin{equation}\label{ktau}
\kappa_\tau \equiv \frac{Y_\tau}{Y_\tau^{\mathrm{SM}}} = \frac{c_\alpha}{s_\beta} + \mathcal{O}\left(\frac{m_\mu}{m_\tau}\right) \times \frac{t_\beta}{s_\beta^2} c_{\beta-\alpha}.
\end{equation}
The order of magnitude terms in the equations above can be calculated by rotating into the flavor basis~\cite{Altmannshofer:2016zrn,Altmannshofer:2018bch}
\begin{equation}\label{muu}
    m'_{\mu \mu }=m_{\mu }+\mathcal{O}\left(\frac{m_{\mu }^2}{m_{\tau }}\right),
\end{equation}

\begin{equation}\label{mcc}
    m'_{cc}=m_c+\mathcal{O}\left(\frac{m_c^2}{m_t}\right),
\end{equation}

\begin{equation}\label{mss}
    m'_{ss}=m_s-m'_{bs}V^\ast _{ts}\left(1+O\left(\frac{m_s}{m_b}\right)\right),
\end{equation}
where $V^\ast_{ts}$ in the above equation comes from the CKM matrix.

The equations above correspond to what we will call the standard
flavorful model, Type 1B.  In this model, the Yukawa textures
$\lambda$ give the couplings to $\Phi_2$ while the textures
$\lambda'$ give the couplings to $\Phi_1$.
The variants described at the end of the
previous section can also be defined. These are, in analogy to the
models defined in that section, the Type 2B model, the Flipped B
model, and the Lepton-specific B model.  The variations are somewhat
intricate to describe in prose, but their patterns of couplings are
given 
precisely in Table~\ref{tab:assignments}.

From the above equations, we can now examine the behavior of these
models  in much the same way as in \cite{Altmannshofer:2016zrn}.  Here,
though, we will shift our focus from the predictions for the heavy
Higgs bosons to the predictions for the 125~GeV boson $h$, for which
the deviations from the SM would be apparent as the result of
precision measurements.

\subsection{Constraints on $\cos(\alpha-\beta)$}
\label{sec:constraints1}

\indent The parameters $\beta$ and $\alpha$ largely determine the
behavior of these couplings, so it is essential to understand how they
behave, and how they are constrained. It is conventional to express
these constraints on plots of $\cos(\alpha-\beta)$ versus $\tan\beta$,
and we will present those plots in this section.   The results for the
conventional and flavorful models -- in our notation, the A and B
models -- are very similar, since these mainly reflect the constraints
on the $h$  couplings to $W$ and $Z$.   As we have noted above,
in the decoupling limit of these models, $\cos(\alpha-\beta)$ is
predicted to be very small, and so the expectation is that the limits
on this parameter will become tighter at higher levels of precision.
This feature distinguishes the 2HDM models from beyond-SM models of
other types, which predict sizable deviations in the $h$ coupling to
$W$ and $Z$.

The current and future constraints on these
parameters are determined by the experimental measurements of Higgs signal strengths.   We follow \cite{Altmannshofer:2018bch} in expressing these through a $\chi^2$ function
\begin{multline}
    \chi^2 = \sum_{i,j} \left( \frac{(\sigma \times \mathrm{BR})_i^{\mathrm{exp}}}{(\sigma \times \mathrm{BR})_i^{\mathrm{SM}}} - \frac{(\sigma \times \mathrm{BR})_i^{\mathrm{BSM}}}{(\sigma \times \mathrm{BR})_i^{\mathrm{SM}}} \right) \\  \left( \frac{(\sigma \times \mathrm{BR})_j^{\mathrm{exp}}}{(\sigma \times \mathrm{BR})_j^{\mathrm{SM}}} - \frac{(\sigma \times \mathrm{BR})_j^{\mathrm{BSM}}}{(\sigma \times \mathrm{BR})_j^{\mathrm{SM}}} \right)(\mathrm{cov})_{ij}^{-1}
\end{multline}
where $(\sigma \times \text{BR})_i^{\text{exp}}$, $(\sigma \times
\text{BR})_i^{\text{SM}}$, and $(\sigma \times
\text{BR})_i^{\text{BSM}}$ stand for the experimental signal strength
results, SM signal strength results and the expected results in
the various 2HDM models~\cite{Altmannshofer:2018bch}. The production mechanisms utilized in this $\chi^2$ function (for the LCF curves) are a combination of vector boson fusion and Higgs-strahlung expected signal strength results primarily for $WW, \gamma\gamma, \tau^+\tau^-, b\bar{b}, c\bar{c}$, and a couple of others. To compare to
the results presented in \cite{Altmannshofer:2018bch}, we use the measured values of $\sigma\times BR$
from  ATLAS
\cite{ATLAS:2019nkf} and
CMS \cite{CMS:2018uag}.  We compare those
results to the expected measurements of
the Higgs signal strengths for the LCF  at $\sqrt{s}=$ 250 GeV, 550
GeV, and 1 TeV with corresponding integrated luminosities of 3, 4 and
8~$\text{ab}^{-1}$,  from \cite{LinearColliderVision:2025hlt}. 
 Following the procedure of  \cite{Altmannshofer:2016zrn} and
\cite{Altmannshofer:2018bch} for the B models,  we determine the order
of magnitude terms in eqs. \leqn{kmu} to \leqn{ktau} by minimizing the
$\chi^2$ function, allowing these numbers to float between 
$\pm 3({m_{\mathrm{2nd}}}/{m_{\mathrm{3rd}}})$. The freedom for this is justified in the theory because of the $\mathcal{O}(1)\times m_{\mathrm{2nd}}/{m_{\mathrm{3rd}}}$ coefficients in the coupling equations.

\begin{figure*}
\centering
\includegraphics[width=.55\linewidth]{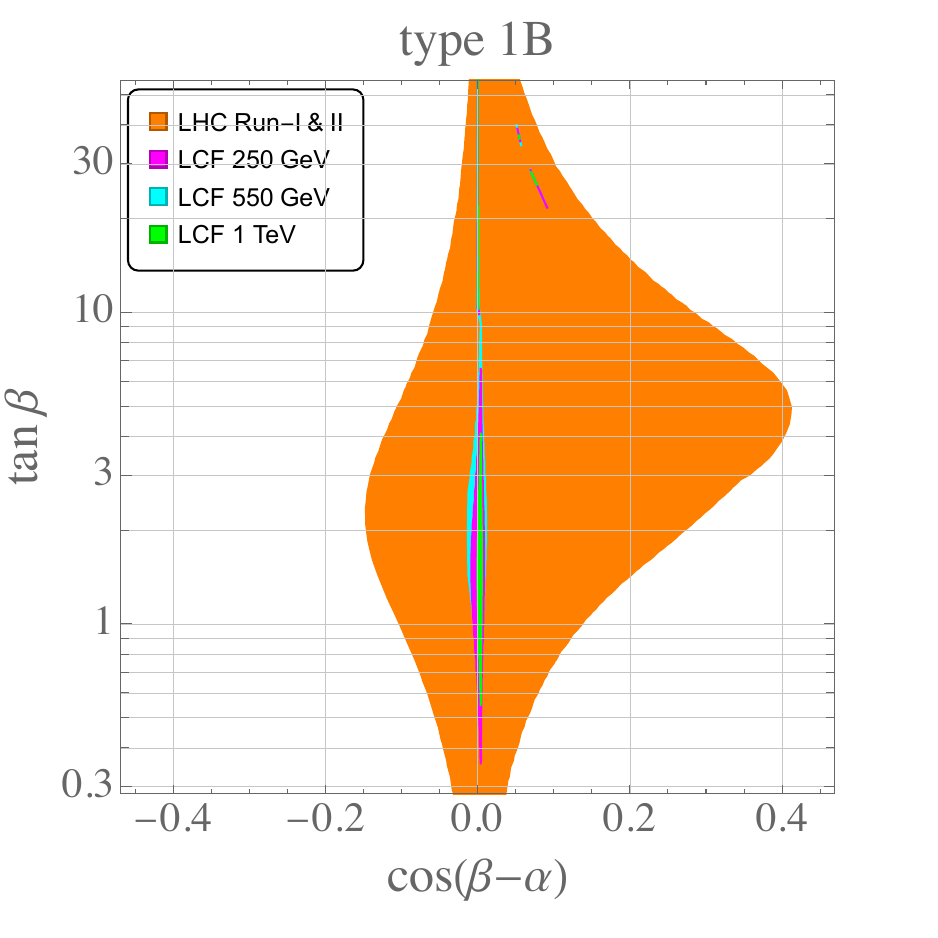}
\caption{The 2$\sigma$ allowed region in the $\text{cos}(\beta-\alpha)$ vs.~$\text{tan}\,\beta$ plane for the type 1B 2HDM using data from ATLAS, CMS and the expected signal strengths for the LCF. Please see the text for more details of the analysis.}
\label{fig:1}
\end{figure*}

\begin{figure*}
\includegraphics[width=.55\linewidth]{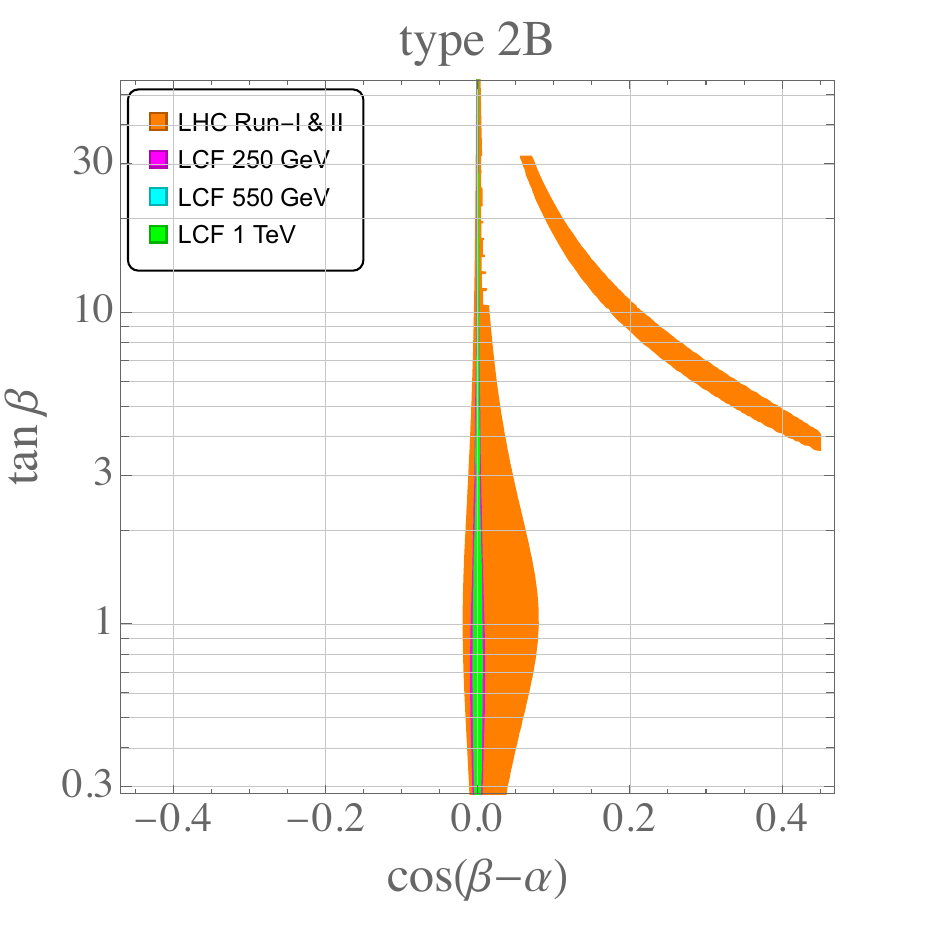}
\caption{The 2$\sigma$ allowed region in the $\text{cos}(\beta-\alpha)$ vs. $\text{tan}\,\beta$ plane for the type 2B 2HDM using data from ATLAS, CMS and the expected signal strengths for the LCF.  Please see the text for more details of the analysis.}
\label{fig:2}
\end{figure*}

\begin{figure*}
\includegraphics[width=.55\linewidth]{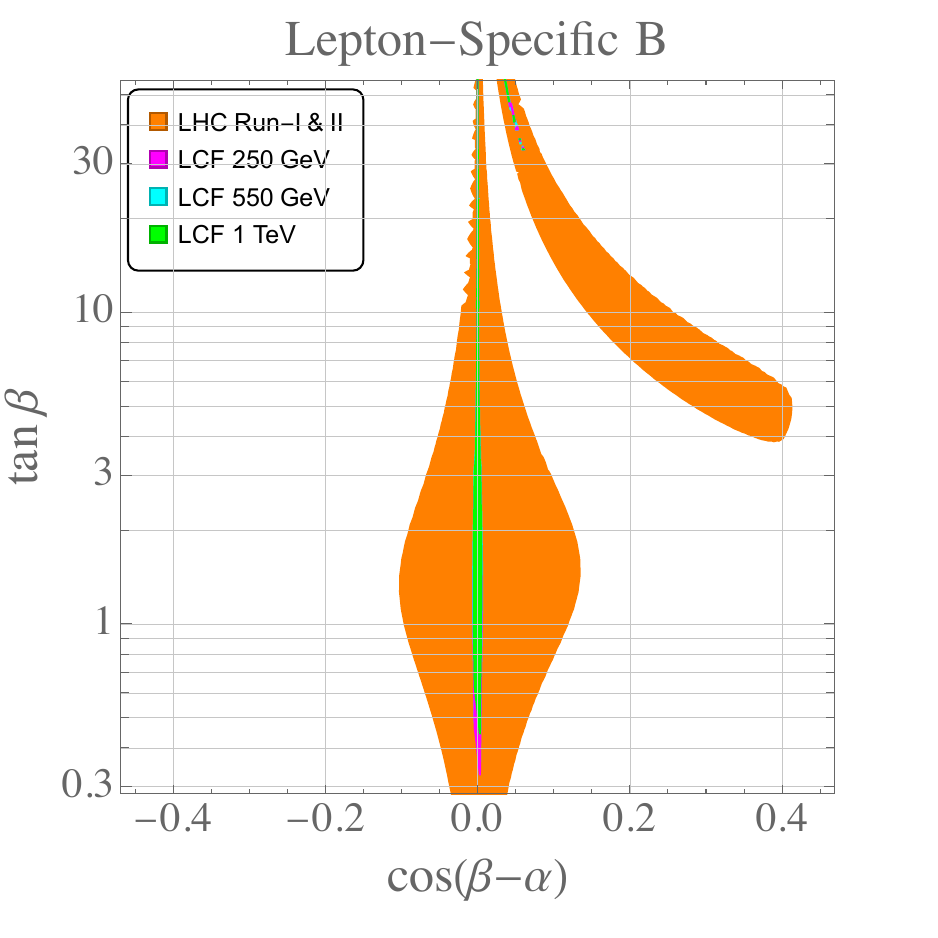}
\caption{The 2$\sigma$ allowed region in the $\text{cos}(\beta-\alpha)$ vs. $\text{tan}\,\beta$ plane for the lepton-specific B 2HDM using data from ATLAS, CMS and the expected signal strengths for the LCF. Please see the text for more details of the analysis.}\label{fig:3}
\end{figure*}

\begin{figure*}
\includegraphics[width=.55\linewidth]{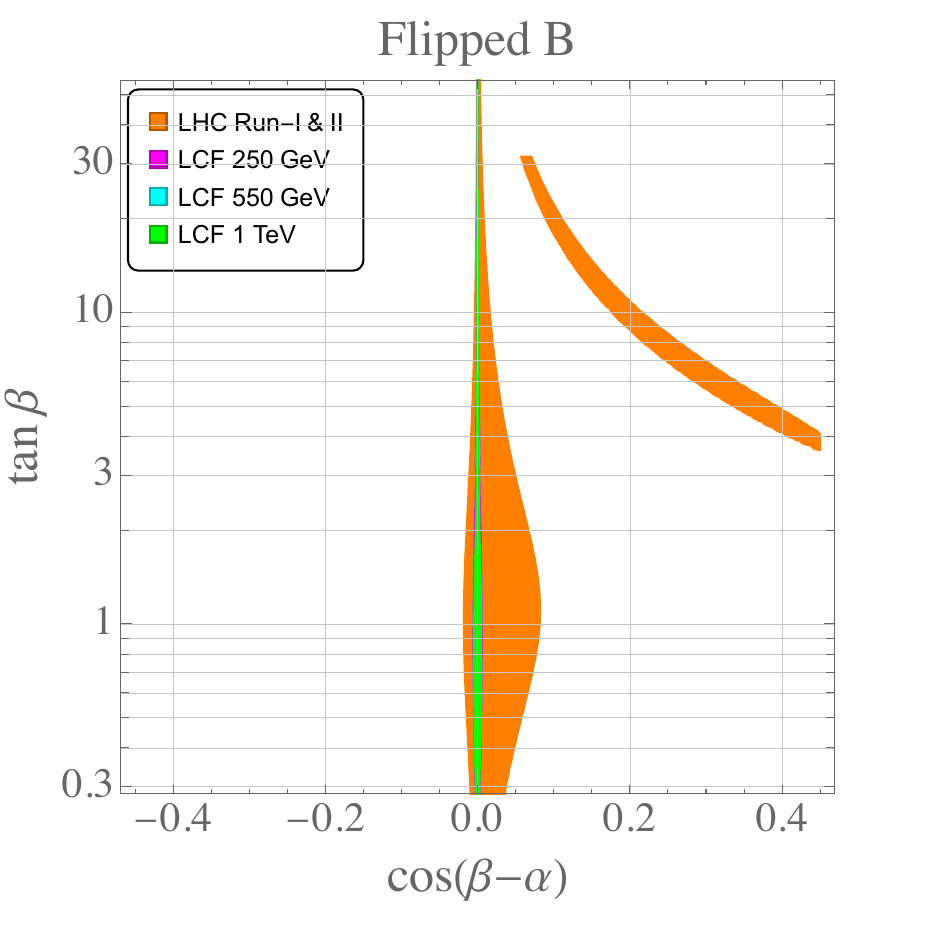}
\caption{The 2$\sigma$ allowed region in the $\text{cos}(\beta-\alpha)$ vs. $\text{tan}\,\beta$ plane for the flipped B 2HDM using data from ATLAS, CMS and the expected signal strengths for the LCF.  Please see the text for more details of the analysis.}
\label{fig:4}
\end{figure*}

The results of this analysis for the B models are shown 
in Figs.~\ref{fig:1}--\ref{fig:4}.  
As expected, the allowed variation of $\cos(\alpha-\beta)$ is already
highly constrained by the current ATLAS and CMS data, corresponding to
the determination of the Higgs boson couplings to $W$ and $Z$ of
better than 10\% in agreement with the Standard Model. It can also be seen, specifically in the Type 1B model, that there are allowed regions corresponding to $\cos(\alpha-\beta)\sim0.1$ and $\tan\beta \sim 30$ that does not appear in the other models. Still, in all cases,
the LCF projected values, which should eventually constrain these couplings to the few
parts per mil level, dramatically contract the allowed region in this
plane.   A similar effect is seen in the corresponding plots for
the A models.


It is interesting that, for all of the B models, the region in which
we see the largest effects of higher precision data is for low values
of $\tan\beta$, between 1 and 5.   This contrasts with the expectation
for the usual Type 2A models, where the primary effects of the 2HDM
nature are at large $\tan\beta$.

\section{Effects on the fermion Yukawa couplings}

\begin{figure*}
\includegraphics[width=.6\linewidth]{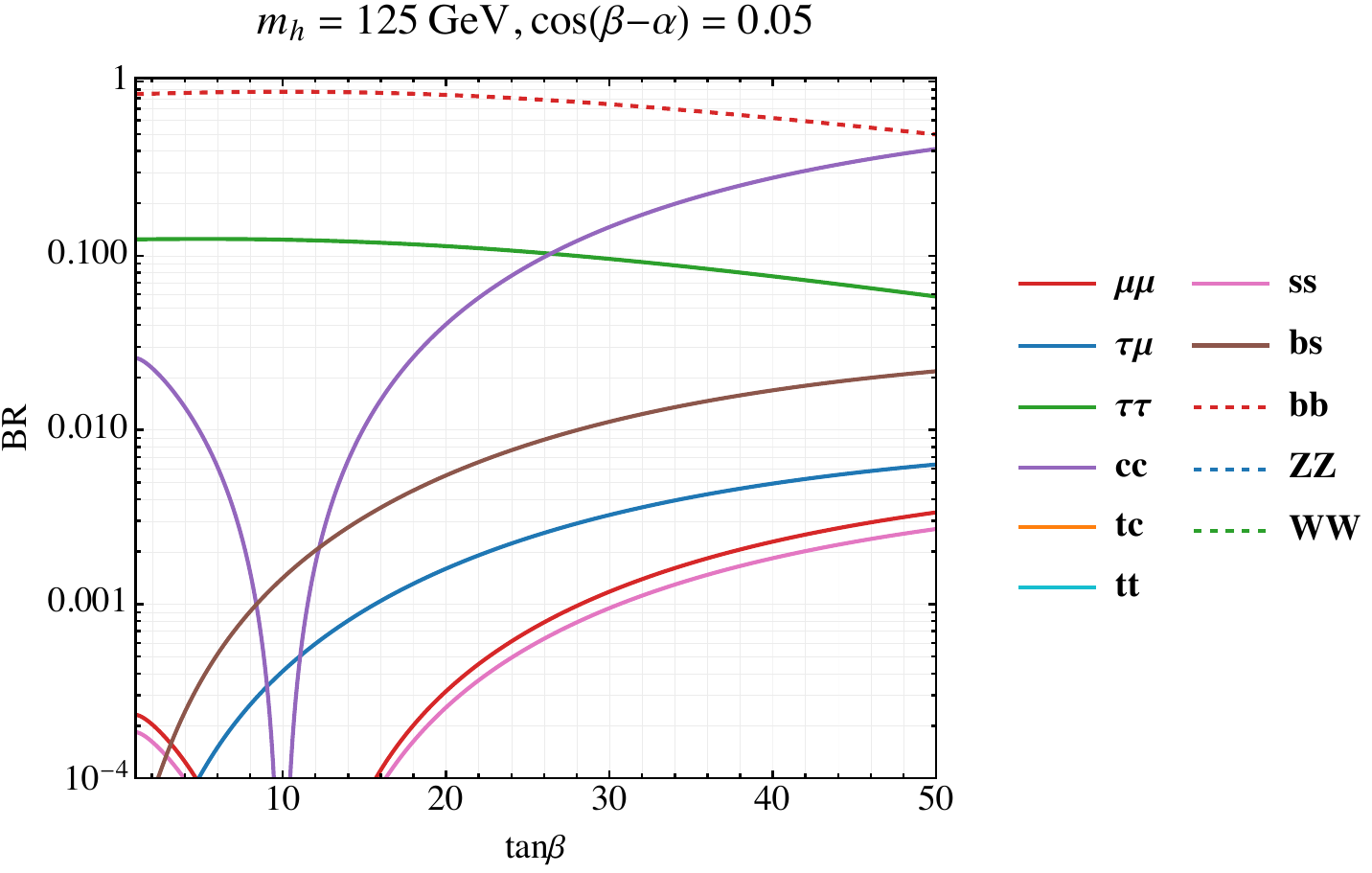}
\caption{The branching ratios for the 125 GeV Higgs in the type 1B 2HDM.}
\label{fig:7}
\end{figure*}

\begin{figure*}[t]
\includegraphics[width=.6\linewidth]{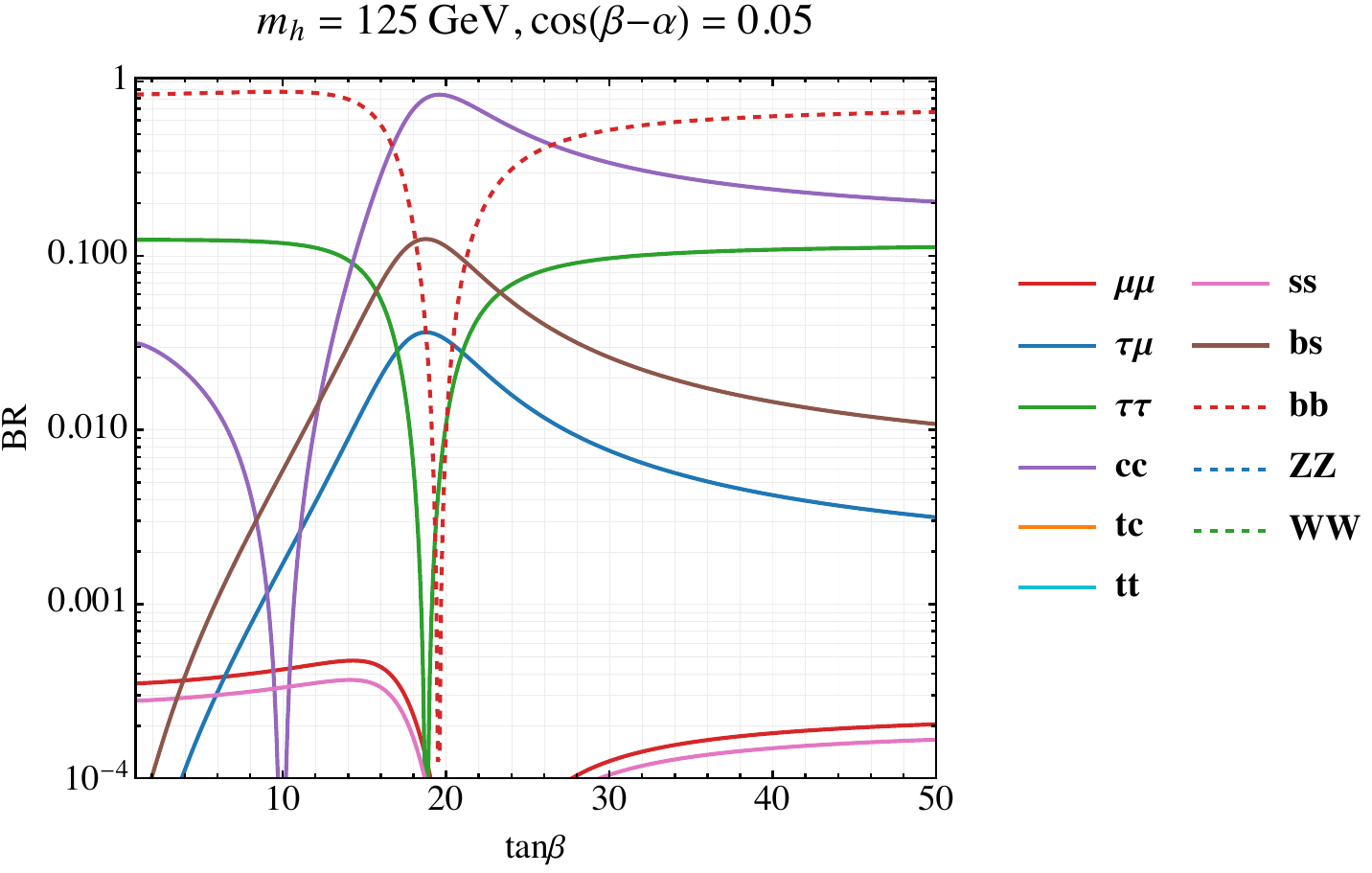}
\caption{The branching ratios for the 125 GeV Higgs in the type 2B 2HDM.}
\label{fig:8}
\end{figure*}

With this preparation, we are now ready to discuss the coupling
modifications expected from the 2HDM in both the A and B cases.  In
the A models, we generally expect an enhancement in the couplings of
the Higgs boson to $b$ or $\tau$ or both.    This effect is also seen
in the B models for the third-generation fermions.  However, in the B
models, the opposite effect is seen for the second-generation
fermions.   The B models also predict flavor-changing Higgs decays such
as $h\to \tau \mu$ and $h\to bs$.   The constraints from CMS
\cite{CMS:2021rsq}
and ATLAS~\cite{ATLAS:2023mvd}, 
\beq
BR(h\to \tau \mu) <  1.5\times 10^{-3} \ , 1.8\times 10^{-3} ,
\eeqn
respectively, at 95\% confidence, restrict us to the low $\tan\beta$
region
$1 < \tan\beta < 10$.  As an illustration of this variation, we show
in Figs.~\ref{fig:7} and \ref{fig:8} 
the branching ratios of a 125~GeV $h$ boson for $\cos(\beta-\alpha) =
0.05$ as $\tan\beta$ is varied.

We are now ready to present the main results of this paper, a survey
of what opportunities for discovery are available from high-precision
measurement of the fermionic Higgs couplings.    For each model, we
characterized the relative coupling deviations predicted in the Yukawa
coupling of the fermion $f$
by a coupling modification $\kappa_f$.  For each model and for each
fermion, we surveyed the possible range of deviations by scanning the
2HDM parameter space.  We carried out the scan as follows: We ignored possible CP violation by assuming that all Lagrangian parameters were real-valued.  Then, using formulae in Appendix D of~\cite{Gunion:2002zf}, we expressed the parameters in the quadratic potential and the quartic coefficients $\lambda_1$-$\lambda_5$ in terms of the measured quantities $m_h^2$ and $v$ and four free parameters,  the potentially observable quantities $m_H^2$, $\tan\beta$, and the quartic coefficients $\lambda_6$, and $\lambda_ 7$, which control the mixing between the two Higgs doublets in this scheme.   We then scanned over these 4 parameters, varying $m_H$ between 1 and 10~TeV, $\tan\beta$ between 1 and 10, and $\lambda_6$ and $\lambda_7$ over the range from 0.1 to 1. This attempts to cover the full range of variation of the 2HDM parameters while still remaining in the perturbative region. When computing the Yukawa coupling deviations, we used the equations \leqn{kt} to \leqn{ktau}.  The input quark masses $m'$ were obtained, after rotating into the mass eigenstate basis,  by taking these to be the  running $\overline{\text{MS}}$ quark masses at the $\mu = 500 \text{ GeV}$ scale, as was done in~\cite{Altmannshofer:2016zrn}.

\begin{figure*}
\includegraphics[width=.8\linewidth]{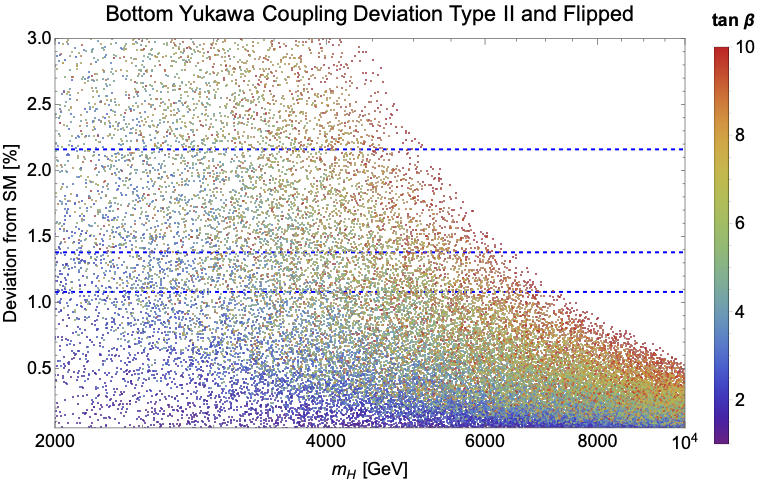}
\caption{A scatter plot for the b quark in the type 2A and Flipped A 2HDM depicting the percent change in the Yukawa coupling from the standard model values created by varying the heavy Higgs mass, $\text{tan}\,\beta$, $\lambda_6$, and $\lambda_7$ from the 2HDM potential. The dashed lines correspond to 3$\sigma$ uncertainties based on the expectations for the  successive LCF energy runs \cite{LinearColliderVision:2025hlt}.}
\label{fig:10}
\end{figure*}

\begin{figure*}
\includegraphics[width=.8\linewidth]{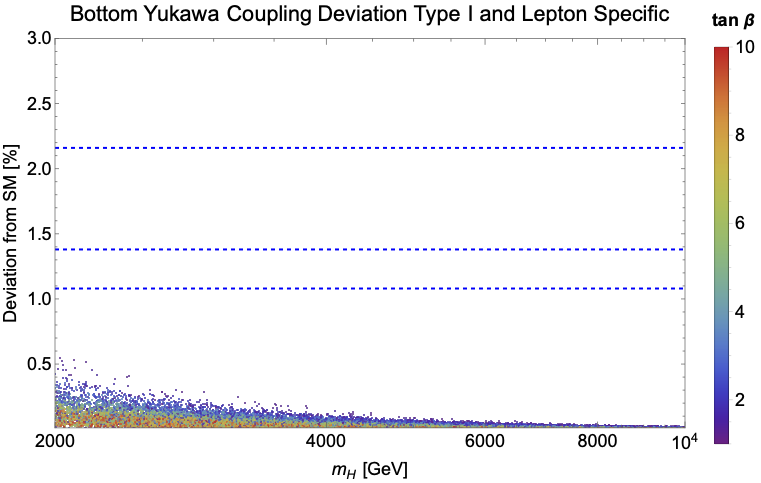}
\caption{A scatter plot for the b quark in the type 1A and Lepton Specific A 2HDM depicting the percent change in the Yukawa coupling from the standard model values created by varying the heavy Higgs mass, $\text{tan}\,\beta$, $\lambda_6$, and $\lambda_7$ from the 2HDM potential.The dashed lines correspond to 3$\sigma$ uncertainties based on the expectations for the  successive LCF energy runs \cite{LinearColliderVision:2025hlt}.}
\label{fig:9}
\end{figure*}

\begin{figure*}
\includegraphics[width=.8\linewidth]{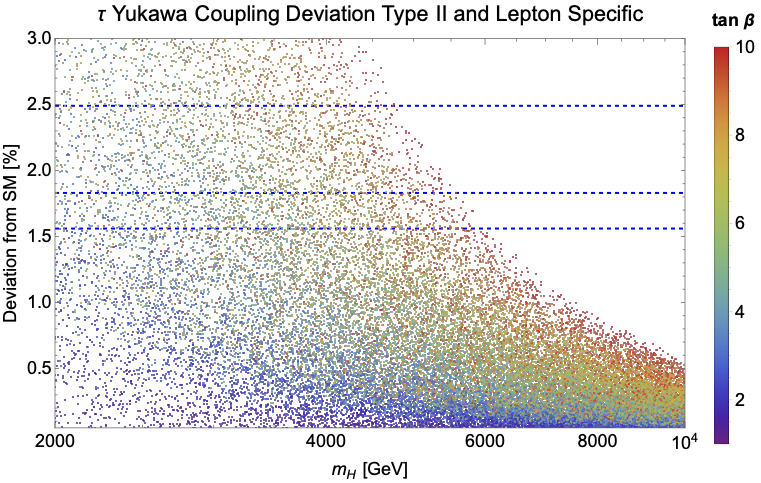}
\caption{A scatter plot for the tau lepton in the type 2A and Lepton Specific A 2HDM depicting the percent change in the Yukawa coupling from the standard model values created by varying the heavy Higgs mass, $\text{tan}\,\beta$, $\lambda_6$, and $\lambda_7$ from the 2HDM potential. The dashed lines correspond to 3$\sigma$ uncertainties based on the expectations for the  successive LCF energy runs \cite{LinearColliderVision:2025hlt}.}
\label{fig:12}
\end{figure*}

\begin{figure*}
\includegraphics[width=.8\linewidth]{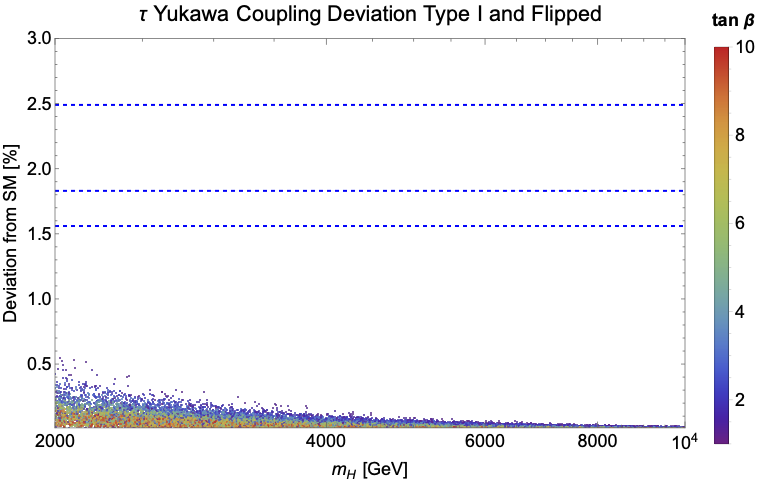}
\caption{A scatter plot for the tau lepton in the type 1A and Flipped A 2HDM depicting the percent change in the Yukawa coupling from the standard model values created by varying the heavy Higgs mass, $\text{tan}\,\beta$, $\lambda_6$, and $\lambda_7$ from the 2HDM potential. The dashed lines correspond to 3$\sigma$ uncertainties based on the expectations for the  successive LCF energy runs \cite{LinearColliderVision:2025hlt}.}
\label{fig:11}
\end{figure*}

\begin{figure*}
    \centering
\includegraphics[width=.8\linewidth]{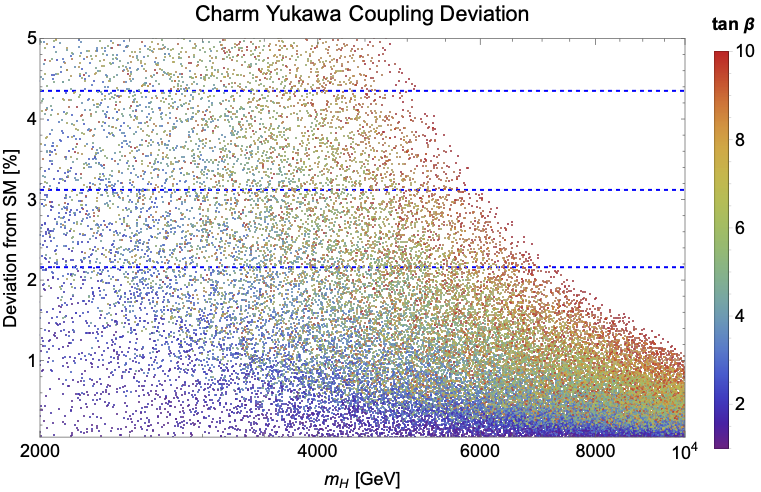}
\caption{A scatter plot for the charm quark in all four flavorful (B) 2HDM depicting the percent change in the Yukawa coupling from the standard model values created by varying the heavy Higgs mass, $\text{tan}\,\beta$, $\lambda_6$, and $\lambda_7$ from the 2HDM potential.The dashed lines correspond to 3$\sigma$ uncertainties based on the expectations for the  successive LCF energy runs \cite{LinearColliderVision:2025hlt}.}
\label{fig:13}
\end{figure*}

\begin{figure*}
\includegraphics[width=.8\linewidth]{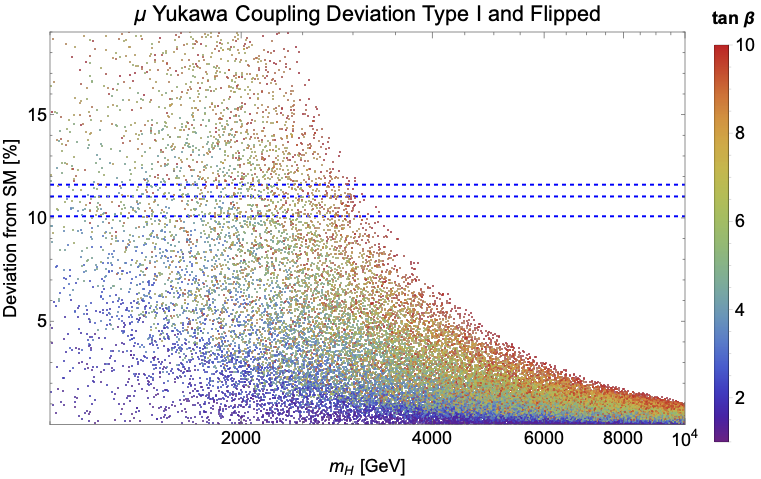}
\caption{A scatter plot for $\mu$ in the type 1B and Flipped B 2HDM depicting the percent change in the Yukawa coupling from the standard model values created by varying the heavy Higgs mass, $\text{tan}\,\beta$, $\lambda_6$, and $\lambda_7$ from the 2HDM potential.The dashed lines correspond to 3$\sigma$ uncertainties based on the expectations for the  successive LCF energy runs \cite{LinearColliderVision:2025hlt}.}
\label{fig:14}
\end{figure*}

\begin{figure*}
\includegraphics[width=.8\linewidth]{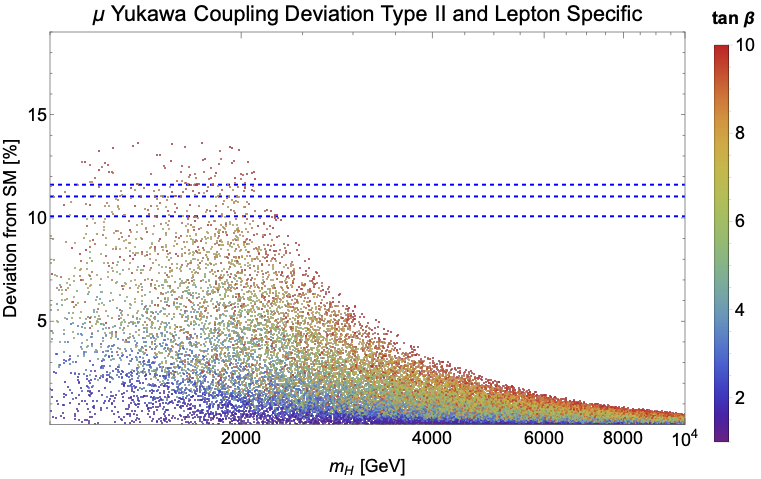}
\caption{A scatter plot for  $\mu$ in the type 2B and Lepton Specific B 2HDM depicting the percent change in the Yukawa coupling from the standard model values created by varying the heavy Higgs mass, $\text{tan}\,\beta$, $\lambda_6$, and $\lambda_7$ from the 2HDM potential. The dashed lines correspond to 3$\sigma$ uncertainties based on the expectations for the  successive LCF energy runs \cite{LinearColliderVision:2025hlt}.}
\label{fig:15}
\end{figure*}
Consider first the A models.   In Figs.~\ref{fig:10} and \ref{fig:9},
we show the results of our parameter scan for the $b$ quark Yukawa
coupling, first in the Type 2A and Flipped A models, and then in the
Type 1A and Lepton-specific 1A models.   The notation is the
following: Points are plotted according to the mass of the heavy $H$
boson.   In the decoupling limit of large $m_H$, the masses of the
other heavy Higgs bosons $A$ and $H^\pm$ are close to the mass of the
$H$.  The vertical scale shows the deviation of the $b$ quark Yukawa
coupling from the SM prediction, in percent.   The three horizontal
lines represent the 3$\sigma$ uncertainty expectations for the LCF at
three stages: after  250~GeV, 3~ab$^{-1}$, after 550~GeV, 4~ab$^{-1}$,
and after 1~TeV, 8~ab$^{-1}$, from \cite{LinearColliderVision:2025hlt}.
The alternative LCF scenario, with  250~GeV, 3~ab$^{-1}$ and 550~GeV,
8~ab$^{-1}$ gives results close to the lowest of these lines.
The expectations for these coupling measurements from the
FCC-ee~\cite{FCC:2025lpp}
and
CEPC~\cite{CEPCPhysicsStudyGroup:2022uwl}
proposals are also similar to the lowest lines shown.
The scan points are color-coded by the value of $\tan\beta$, from 1
(blue) to 10 (red).

The obvious cutoff with increasing $H$ boson
mass reflects the $1/m_H^2$ decoupling of the heavy Higgs multiplet.
The density of points in the plots has no significance; roughly, an
equal number of points were thrown in each vertical slice, and so
these have higher or lower density according to the height of the column.

The sensitivity of the $b$ Yukawa coupling is clearly different in the
different coupling schemes.   This is to be expected. In the 1A model, the expected signals are at most a few parts per mil, well below the 3$\sigma$ criterion for any Higgs factory.  What is more
surprising is the sensitivity to very large $H$ boson masses in the
case of Fig.~\ref{fig:10}.   There are models with observable
deviations with $H$ masses above 5~TeV.    The models giving relatively large
deviations are, in general, those with stronger quartic couplings.
Often, surveys of the parameter space of 2HDM models assume a
potential compatible with supersymmetry.  As was pointed out in
\cite{Peskin:2022pfv}, this assumption not only constrains the coupling scheme in
2HDM models but it also constrains the strength of the nonlinear
couplings, which arise from D terms and thus are related to electroweak
gauge couplings.  Exploring the full parameter space of 2HDM models
indicates much more opportunity for the discovery of heavy Higgs boson
effects through Higgs precision measurements.

The analogous plots for the tau lepton Yukawa coupling are shown in
Figs.~\ref{fig:12} and \ref{fig:11}.   Note that, in this case, it is
the Lepton-specific model that offers exceptional sensitivity.

We now turn to the flavorful B models.  Here, the large effects on
Yukawa couplings appear for the second generation.  These are arranged
so that the charm quark has the largest modifications in all four
models.   The scan of parameter space is shown in Fig.~\ref{fig:13}.
It is important to note that the measurement of the charm quark Yukawa
will give the largest improvement from LHC Yukawa coupling
measurements to the Higgs factory measurements, from order-1
constraints to constraints at the sub-\%\ level.   The discovery of
large deviations from the SM predictions specifically in the charm
quark Yukawa coupling would be a remarkable verification of the idea
that the fermion mass spectrum is generated by an underlying
generation-dependent Higgs sector.   From the figure, this opportunity
will be available in the 2HDM parameter space even for some models
with $H$ boson
masses above 5~TeV.  

The expected constraints from measurements of
the muon Yukawa coupling, either from the HL-LHC or from Higgs
factories, are expected to be significantly weaker, at about the 3\%
level.   Nevertheless, this adds an opportunity for the confirmation
of the flavorful Higgs idea.    The plots for the deviations in the
muon Yukawa couplings in the different coupling scenarios are shown in
Figs.~\ref{fig:14} and \ref{fig:15}.

\section{Conclusions}

In this paper, we have discussed the implications of models with
extended Higgs sectors on the measurable Yukawa couplings of the
125~GeV Higgs boson.    The sensitivity of Yukawa coupling
measurements expected at Higgs factories depends strongly on the
particular scenario by which the heavy Higgs bosons couple to the
various quark and lepton flavors and generations.   However, we have
shown that, in these models, the sensitivity to large values of the heavy Higgs boson
masses is much larger than is commonly appreciated, with sensitivity
to masses of 5~TeV and above in significant regions of the parameter
space.  This is an opportunity to discover new physics relevant to the
mystery of the fermion mass hierarchy that should not be missed.

We find it especially interesting that this conclusion extends to the
``Flavorful
2HDM''
model~\cite{Altmannshofer:2016zrn,Altmannshofer:2017uvs,Altmannshofer:2018bch}
and its generalizations.   In this model, the fermion mass hierarchy
arises from the fact that the underlying structure contains two Higgs
boson doublets that couple differently to the various generations.
The discovery of a large deviation from the SM prediction specifically
in the charm
Yukawa coupling would provide strong evidence for this idea.  This
discovery is very much possible at Higgs factories and actually is a
primary motivation for making these measurements.

More generally, the pattern of deviations in the Yukawa couplings of
the various quarks and leptons -- $t$, $b$, $\tau$, $c$, $s$, and
$\mu$ -- encode information about the details of the new physics
present at higher energies.   The studies in this paper provide
concrete examples of this more general philosophy, illustrated more
broadly, for example, in \cite{Barklow:2017suo}.

Although one can estimate the effects of heavy new particles on future
precision measurements of the parameters of the SM, either by
order-of-magnitude estimation or from Effective Field Theory, there is
no replacement for carrying out explicit parameter scans in specific
UV-complete extensions of the SM.   We look forward to further
exploration of the opportunities that precision Higgs physics will
make available.

\begin{acknowledgments}
We thank Douglas Tuckler for his valuable conversations and guidance as we tried
 to replicate and build upon his previous work. We also thank Wolfgang Altmannshofer for his comments on the manuscript. The work of
KM and DGEW
was supported in part by the grant NSF OIA-2033382.   The work of MEP was
supported by the US Department of Energy under contract DE–AC02–76SF00515.
 \end{acknowledgments}

\bibliography{Bibliography2}

\end{document}